\documentclass[runningheads,citeauthoryear]{apinv}
\usepackage{epsfig,cite,graphics}
\usepackage{marvosym} %For astronomical symbols % % %
\usepackage[utf8]{inputenc}

\begin{document}

\title{Post - superhumps maximum on intranight time scales of the AM CVn star CR Boo }
\author{D.Boneva\inst{1}, G.Latev\inst{2}, S.Boeva\inst{2}, K.Yankova\inst{1}, R.Zamanov\inst{2}   }
\titlerunning{Intranight superhumps of CR Boo}
\authorrunning{D.Boneva, et al.}
\tocauthor{G.Latev, S.Boeva,\inst{2}}
\tocauthor{K.Yankova,\inst{1}}
\tocauthor	{R.Zamanov, \inst{2}} 
% Command tocautor{} is used by the Latex to give author names 
% to the Contents of the volume (automatically generated)
\institute{Space Research and Technology Institute, Bulgarian Academy of Sciences, BG-1113, Sofia
		\and Institute of Astronomy and NAO, Bulgarian Academy of Sciences, BG-1784, Sofia   \newline
	\email{danvasan@space.bas.bg}    }
\papertype{Submitted on 12.04.20123; Accepted on 12.05.2023}	
% Papertype can be "Research report", "Review", "Invited lecture", "Conference talk", 
% "Conference poster", "Lecture at scientific seminar", "Summary of dissertation",  etc.
\maketitle

\begin{abstract}
We present observations of the intranight brightness variability of CR Boo, a member of the AM CVn stars group. The observational data are obtained with the 2m telescope of the Rozhen National Astronomical Observatory and the 60 cm telescope of the Belogradchik Observatory, Bulgaria, in BVR bands. We report the appearance of superhumps, with an amplitude from 0.08 to 0.25 mag, when the maximum brightness reaches the magnitude 14.08 in the V band, and 14.13 in the B band. A secondary maximum of each superhump is detected with the same periodicity as the superhumps: $Psh = 24.76 - 24.92$ min. In our results, the post maxima are shifted in time from $\approx 7.62$ min to $\approx 16.35$ min in different nights, with an amplitude of $\approx 0.06 - 0.09$ mag and an amplitude difference of $\approx 0.035$ mag towards the superhumps’ maximum. We find a correlation of the post maxima with the accretion processes at the outer side of the disc. 
      
\end{abstract}

\section{Introduction}

AM CVn stars are double white dwarf binaries, initially possible to be detected only by their helium emission lines (Wood et al 1987, Provencal et al 1997, Patterson et al 1997, Kato et al. 2000). In the paper of Ramsay et al. (2018), the number of $\approx 56$ known AM CVn stars is reported. They are binaries with short orbital periods, of 5 – 65 minutes (Podsiadlowski et al. 2003, Solheim 2010).
AM CVns are part of the cataclysmic variables (CVs) stars family (Warner 1995). The most possible evolutionary channels schemes of AM CVns include a helium-reach donor star, in contrast with the hydrogen-rich secondary component in the ordinary CVs.  
A low frequency gravitational wave (GW) radiation is detected in AM CVn stars by eLISA  (Evolved Laser Interferometer Space Antenna) (Nelemans 2013). The GW emission has an important role in evolving the binary into their semidetached phase (Solheim 2010). One main feature of AM CVns is that the white dwarf accretes from another white-dwarf companion (Nelemans et al. 2001, Paczyñski 1967, Faulkner et al. 1972), where the donor star could be semi- or fully-degenerated. 
Further mass-transfer between the components, by its stability or instability factor, has a significant effect on the evolution of the white-dwarf binary configuration in AM CVn stars (Marsh et al. 2004). For the binaries with sufficiently short orbital periods, the rate of angular momentum loss is efficient through the gravitational wave emission (Paczy\'{n}ski 1967, Faulkner et al. 1972)

As a member of the AM CVn stars group, CR Boo is an interacting double white dwarf binary (Paczy\'{n}ski 1967, Faulkner et al. 1972, Kato et al. 2000, Nelemans et al. 2004), discovered in 1986 by Palomar - Green Survey (Green et al. 1986). 
From the first observations of this objects (Wood et al. 1987) and up to now, its brightness varies with amplitude 13.0 - 18.0 mag in the V band. 
In the CR Boo’s spectrum, He I lines are observed (Wood et al. 1987). 
The orbital period of CR Boo is estimated as Porb = 0.017 days (Provencal et al. 1997, Isogai et al. 2016), which is about 24.5 min or 1471.3 s. Judging by this orbital period, CR Boo belongs to a group characterized with regular outbursts or occasional super-outbursts production, and with a variable size of the disc (Solheim 2010).
As an outburst system (Kato et al. 2000, Groot et al. 2001), CR Boo periodically passes from faint to bright states and it manifests a brightness variability in a range of 1-3 magnitudes at optical wavelengths (Isogai et al. 2016, Duffy et al. 2021, Boneva et. al. 2020, 2022). The object shows characteristics of SU UMa type dwarf novae (Patterson et al. 1997, Kato et al. 1999), which exhibit short-time normal outbursts and longer superoutbursts that can last for weeks.  During the outbursts periods of CR Boo, a production of superhumps is observed (Isogai et al. 2016, Boneva et al. 2022). Superhumps are short-period, low-magnitude brightness variations, and when they are positive, their periodicity is a few percent longer than the binary period (Kato et al. 2000, Patterson 2005). Superhumps can be observed during the outbursts state of the cataclysmic variables and AM CVn stars (Warner 1995).
Currently, we construct phase-average diagrams, based on the superhumps periodicity (Sections 3.1 and 3.2). The appearance and discussion on the post-superhumps maxima are presented in Sections 3.2 and 4.

\section{Observations and data reduction}

We report the observational data, obtained with the 2.0 m telescope of the National Astronomical Observatory (NAO) Rozhen, Bulgaria and 60 cm telescope of the Belogradchik Observatory, Bulgaria.
The 2m telescope with CCD camera Andor iKON-L, with 2 – channel focal reductor FoReRo was used on February 12, 2021 in the BVR bands. The observations on April 16th 2020 in B band were obtained with the 60 cm telescope of the Belogradchik Observatory (with CCD camera FLI PL16803). 
 
Data reduction was performed with standard tools for processing of CCD images and aperture photometry. The photometric standards were applied. Six comparison stars have been used, based on the standards in the APASS9 catalog with their original data (see table in Boneva et al. 2022). 

A periodicity of the maximum brightness were obtained by PDM (Phase Dispersion Minimization) method by Stellingwerf (1978). We used the PGRAM (https://exoplanetarchive.ipac.caltech.edu) and PerSea (Maciejewski \& Niedzielski, 2005) software packages to check the results.

\section{Observational results}

\subsection{Superhumps periods of CR Boo}

 We present our observational results of CR Boo for two nights, obtained in different campaigns: 16 April, 2020 (hereafter 20200416) and 12 February, 2021 (hereafter 20210212). In both nights CR Boo was in an outburst state (see Boneva et al. 2022).
 On the night of 20200416, the average magnitude of the star was $14.06 \pm 0.02$ in the B band. We have detected an appearance of superhumps on this date (Fig. 1a), with periodicity of $Psh \approx 24.76 \pm 0.023$ min, estimated in Boneva et al. (2022).  
 On the second night, 20210212, the magnitude was $14.17 \pm 0.01$ in B and $14.22 \pm 0.01$ in R (Boneva et al. 2022). Here, the light curve is made for a shortened time-period with a more detailed view, where the superhumps during this night are clearly distinguished (Fig. 1b). Additionally, data in the V band are also included. We estimated the superhumps periodicity on this date as $Psh \approx 24.92 \pm 0.0012$ min in Boneva et al. (2022).       
 
 \begin{figure}[!htb]
  \begin{center}
    \centering{\epsfig{file=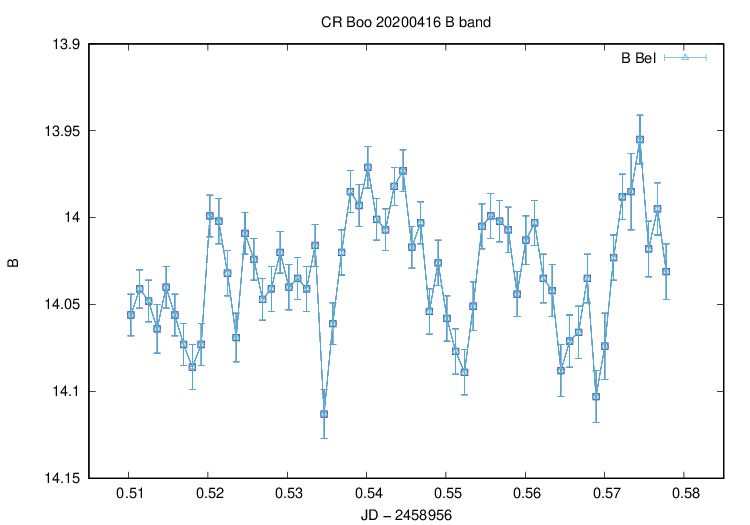, width=0.45\textwidth}}
    \centering{\epsfig{file=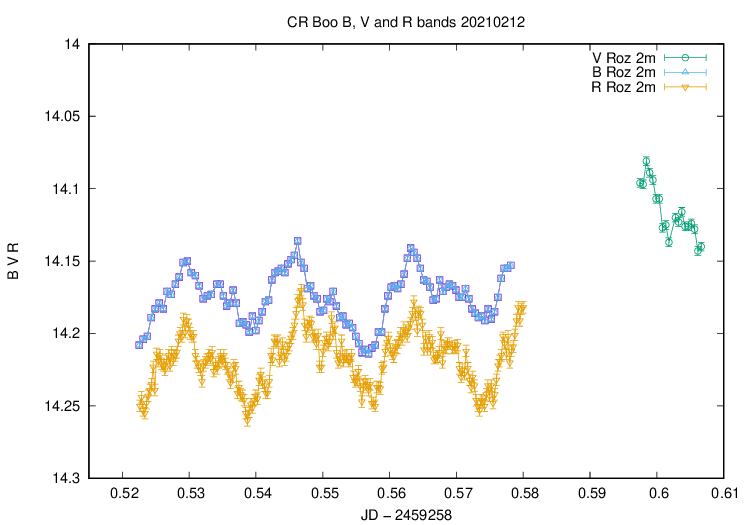, width=0.45\textwidth}}
    \caption[]{Light curve of CR Boo’s intranight observations in B band (left - a) and BVR bands (right - b). Superhumps are detected in both nights. The data are obtained with the 60 cm telescope of the Belogradchik Observatory, Bulgaria and the 2m telescope of NAO Rozhen.  
    } 
   \label{fig:1}
  \end{center}
\end{figure}

\subsection{Phase – average diagrams and post-superhumps maxima}

 During the superhumps, the observations show secondly lower maxima of the brightness, calling them post - superhumps maximum or shortly post - superhumps, which appeared with a period similar to the superhumps periodicity. It is difficult to see some secondary maxima on the light curve during the night of 20200416. They are well distinguishable at the phase-average diagram, obtained on this night (Fig. 2). We estimate their shift from the superhumps maxima as $\approx 16.35 \pm 0.05$ min, with amplitude $0.098 \pm 0.012$ mag against the brightness minimum.
 Such post-superhumps are observed at the phase average - magnitude diagram (Fig. 3), constructed with the data from 20210212. Their average time-shift is $\approx 7.62 \pm 0.005$ min with an amplitude difference towards the superhumps' maximum $\approx 0.035$ mag. These post-superhumps maxima appear at interval similar to the superhump’s period $\approx Psh = 24.92$ min to $\approx 25.03$ min in the B and R bands. They can be seen on the light curve of this night as well (Fig. 1b).

  \begin{figure}[!htb]
  	\begin{center}
  		\centering{\epsfig{file=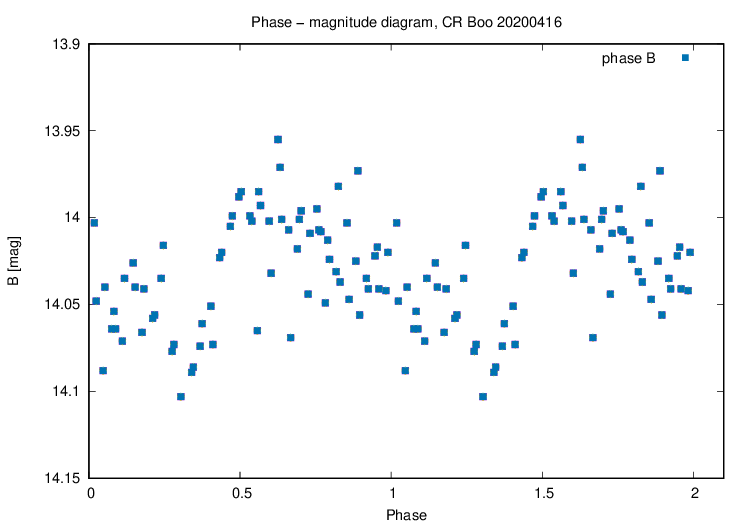, width=0.50\textwidth}}
  		%\centering{\epsfig{file=fig1b.eps, width=0.45\textwidth}}
  		\caption[]{Phase average - magnitude diagram in B, based on the obtained superhumps periodicity $Psh = 24.76 \pm 0.023$ min, on the night of 20200416. } 
  		\label{fig:2}
  	\end{center}
  \end{figure}

\begin{figure}[!htp]
	\begin{center}
		\centering{\epsfig{file=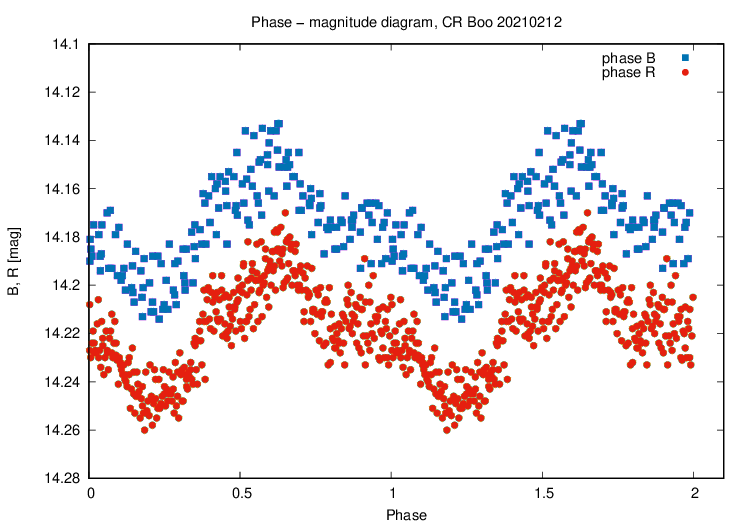, width=0.50\textwidth}}
		%\centering{\epsfig{file=fig1b.eps, width=0.45\textwidth}}
		\caption[]{Phase average – magnitude diagram, constructed on the superhumps periodicity of intranight observations on 20210212.  
		} 
		\label{fig:3}
	\end{center}
\end{figure}

The periodograms in Figs. 2 and 3 are the average result, obtained by analysis of PGRAM, PerSea and PDM (see Section 2).

\section{Discussion}

\subsection{ An intranight post-superhumps detection} 

According to our observational data, it is seen that the small - amplitude modulations in brightness appear with periods close to the orbital period. 

Post-superhumps maxima are possible to be observed during one night, because of the short orbital period of CR Boo. They appear when the star is in an outburst state. In comparison, we haven’t detected any secondary maxima on the night of July 5, 2019, when the star was in a low state and definitely with a humps activity. They are not seen on the intranight light curve (see Boneva et al. 2020, 2022) or on the phase-average diagram of this date (Fig. 4).  

\begin{figure}[!htb]
	\begin{center}
		\centering{\epsfig{file=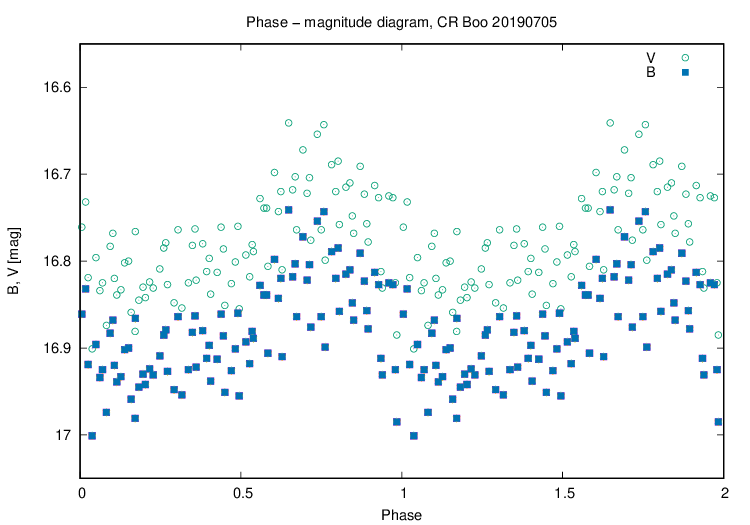, width=0.50\textwidth}}
		%\centering{\epsfig{file=fig1b.eps, width=0.45\textwidth}}
		\caption[]{Phase average – magnitude diagram in the B, V bands, based on the humps periodicity of the intranight observations on 20190705.  
		} 
		\label{fig:4}
	\end{center}
\end{figure}

Superhumps have also been detected in many SU UMa variables. 
The phase average diagrams of V 1047 Aql, GS Cet and NY Her show post-superhumps appearance during their super outbursts in 2016 (Kato et al. 2017).

\subsection{Post-superhumps sources}

In Boneva et al. (2022) we have discussed the possible mechanisms and sources of the superhumps production in CR Boo. 
Several mechanisms could cause an appearance of superhumps, summarizing the most probable of them as: the disc precession; tidal waves; spiral-density formation.
With its superhumps and superoutbursts characteristics, CR Boo is also counted into the SU UMa class of dwarf novae stars (Warner 1995). The superhumps production for these objects is explained by the precession of the eccentric disc that cause some kind of a periodically beating (Whitehurst 1988). The appearance of the post-superhumps maxima could be caused by the switching-on of a second mechanism during the outbursts period, as an example of a second instability. 
It is very likely that at some point of time, a tidal wave inflowing to the primary star's accretion disc in a combination with the precession of such a disc might produce superhumps (Hirose \& Osaki 1990, Wood et al. 2011, Kato et al. 2017).
 
It is known that even a small change in the inflow tidal wave’s velocity could make transformations in the outer accretion disc (Bisikalo et al. 2008, Boneva et al. 2009). 
This usually causes a destabilisation eﬀect on the hot spot structure, as at a further stage, it could turn to a hot line formation. We assume that the hot line might produce the secondary maxima during the superhumps. 
Then, the system could become bright and blue during the superhumps, which, on the other hand, is in contradiction with the results in (Boneva et al. 2022), where we found that the star is bright, but redder on the second date - 20210212.
  
Honeycutt et al. (2013) make a detailed analyses of CR Boo's lightcurves for longer-term observations. They have an interesting suggestion that the star, as a physical system, and its light curves have a chaotic behavior under the rules of the deterministic chaos principles.

 \subsection{On the source's parameters}
  
In this section, we make an analytical estimation of the probable superhump's source "size", as we rely on the temperature and luminosity assumptions for the two observational dates.  We use the Stefan-Boltzmann's law and we adapt the formula for our study:

$L_{s} = \sigma T_{eff}^{4} ( 4 \pi R^{2}_{s} ) $

 Where, we can assume that: $L_{s}$ is the observational luminosity of the source; $T_{eff}$ is the effective temperature of the source in this case; $\sigma$ is the Stefan-Boltzmann constant. We define $R_{s}$ as a size of the superhump source. 

Further, we use the relation between the effective and color $T_{col}$ temperatures: $T_{eff}^{4} = \tau T_{col}^{4}$ - in geometrically thin radiative layers. Where $\tau$ is the optical thickness of the layer.

%We use denotations $L_{s1}$, $T1_{eff}$ and $R_{s1}$ which refer as to the night of April 16, 2020 and $L_{s2}$, $T2_{eff}$ and $R_{s2}$, respectively for night of February 12, 2021.  

 We also denote with $"1"$ the terms which refer as to the night of 20200416 and $"2"$ respectively for the night of 20210212.

Now, we can express a next relation between the luminosities for the two nights, as: 

  $\frac{L_{s1}}{L_{s2}}=\frac{T_{eff1}^4}{T_{eff2}^4}\frac{R_{s1}^2}{R_{s2}^2}=
  \frac{\tau_{1}T_{col1}^4R_{s1}^2}{\tau_{2}T_{col2}^4 R_{s2}^2}$

Then, using the distance to the object, \textbf{$d(CRBoo)=337 pc$} (Sion et al. 2011), and based on its apparent magnitudes, we obtain the relation of the observational luminosity  $L_{s1}/L_{s2} = 0.95 \pm 0.02$. 
For the color temperature, we have $T_{col1}/T_{col2} = 1.218 \pm 0.024$, where the values for the two nights are obtained in (Boneva et al. 2022). Applying these values to the eq.2, gives: 

$\frac{\tau_{1}(\rho_{1})R_{s1}^2}{\tau_{2}(\rho_{2})R_{s2}^2} \approx 0.43 \pm 0.02$

%$(\frac{R_{s1}}{R_{s2}})^2 \approx 0.43 \frac{\tau_{2}\rho_{2}}{\tau_{1}(\rho_{1})}$

The ratio between the sizes $R_{s1}$ and $R_{s2}$ depends on the optical thickness in a function of the mass density, in this way: 

$\frac{R_{s1}}{R_{s2}} \approx \sqrt{0.43 \frac{\tau_{2}(\rho_{2})}{\tau_{1}(\rho_{1})}}$

This leads to the rude estimation that the size $R_{s1}$ in the first night, when CR Boo is bluer is $\approx 0.66 \sqrt{ \frac{\tau_{2}(\rho_{2})}{\tau_{1}(\rho_{1})}}$ times the size $R_{s2}$ in the second night, when the object is redder.

\section {Conclusion}

We presented our observations of intranight brightness variability of the AM CVn star CR Boo, in the BVR bands. We reported an appearance of superhumps, with an amplitude from 0.08 to 0.25 mag and post-superhumps maxima with the same periodicity as the superhumps: $Psh = 24.76 - 24.92$ min. In our results, these secondary maxima in brightness are shifted in time from the primary maxima with $\approx 7.62$ min to $\approx 16.35$ min in different nights. They have an amplitude of $\approx 0.06 - 0.09$ mag and an amplitude difference of $\approx 0.035$ mag towards the superhumps’ maximum. We found the post-superhumps maxima of CR Boo are detected during the periods of outbursts activity in the current observations. This is visible both in the light curves and phase average – magnitude diagrams. We estimated the ratio of the superhumps' size between two nights. A correlation of the post - superhumps maxima with the accretion processes at the outer side of the disc is very possible.

===========

\textbf{Acknowledgments}: This work is supported by the grant "Binary stars with compact objects", $K\Pi -06-H28/2$    $08.12.2018$ (Bulgarian National Science Fund).
D.B. thanks for the support to the EUROWD22 workshop organizers, where part of these results were presented.

\end{document}